# "User Interfaces" and the Social Negotiation of Availability

**Paul M. Aoki and Allison Woodruff**
Palo Alto Research Center
3333 Coyote Hill Road
Palo Alto, CA 94304-1314  USA
aoki@acm.org, woodruff@acm.org

**ABSTRACT**
In current presence or availability systems, the method of presenting a user's state often supposes an instantaneous notion of that state – for example, a visualization is rendered or an inference is made about the potential actions that might be consistent with a user's state.  Drawing on observational research on the use of existing communication technology, we argue (as have others in the past) that determination of availability is often a joint process, and often one that takes the form of a negotiation (whether implicit or explicit).  We briefly describe our current research on applying machine learning to infer degrees of conversational engagement from observed conversational behavior.  Such inferences can be applied to facilitate the implicit negotiation of conversational engagement – in effect, helping users to weave together the act of contact with the act of determining availability.

**INTRODUCTION**
The most common application of presence and availability systems is to advise users when conditions might be suitable for making contact – e.g., facilitating communication between a user who wishes to make contact (hereafter *contactor*) and one or more *contactees*.  Contact can be face-to-face, as in Ambush [8], as well as technologically-mediated, as in Awarenex [5] and BESTCOM [6].  Typically, such systems provide some kind of visible *representation* of the contactees' presence or availability state (past, present, and/or projected) so that the contactor can make appropriate decisions.  For example, BESTCOM infers which communication channels are most appropriate for contactees given their context and presents these choices to the contactor (without revealing the context itself).  Awarenex applies inference to, e.g., automate "away" status messages as an aid to contactors.  Ambush provides a temporal visualization.

Clean, simple representations seem suitable for groups with well-aligned goals and practices – gelled work teams, departments, etc.  They also seem valuable in complex availability scenarios, such as determining good meeting times for groups.  However, different kinds of groups may present additional challenges for availability forecasting.  Consider the differences between co-located work within an organizational workplace (which is likely to have considerable synchrony between individuals' activity rhythms), distributed work within an organizational workplace (in which rhythms may be offset by geographic time differences or rearranged by work practice differences between the sites), and telecommuting and nomadic work (in which rhythms are strongly influenced by constant interaction with external entities and the contingencies of travel).  In situations where availability forecasting may be less reliable, or where social relationships are such that contact may be sensitive, it may be useful to conceptualize our representations of "availability" in a different way.  The idea would be that such representations ought to lead to "socially" robust results in the face of system mispredictions or user misinterpretations.

In what follows, we briefly point out some challenges – "sensitivities" might be a better word – that could be considered in the design of systems that present presence and availability to contactors.  We then describe our current research, which focuses on the use of machine learning to assess the progress of conversational engagement.

**USER INTERFACE CHALLENGES**
We argue that the most effective systems for sharing presence and availability information will reflect naturally occurring social processes.  For example, they will allow information to flow back and forth between the contactee and the contactor.  In this section, we discuss how availability information is part of an ongoing process of communication.  The three main observations are somewhat interrelated, corresponding to "intuitive" dimensions rather than "orthogonal" dimensions.  Each observation raises a number of issues, which we present as challenges rather than as prescribed solutions.  In the following section, we discuss our own approach to these challenges, which is quite different from that taken in the most obviously related systems (e.g., [5,6,8]).

**Expression of availability is highly contextual**
 Representations of a contactee's availability tend to be "one size fits all" with respect to situational context.  Some systems do support, e.g., access control rules which allow contactees to control what information will be presented to what contactors.   Otherwise, contactees generally have little ability to define how they present themselves – to act out different *lines* [4] – to different contactors in different situations.

In the absence of computers, the manner in which contactees express availability is often highly contextual. That is, how contactees choose to express their current state and activities to others depends on their current understanding of the situations and needs of all parties. Schegloff notes that the formulations of *place* that are communicated in telephone conversation – where "place" actually includes broad notions of situational context – depend on who is calling, the purpose of the call, how the callee wishes to present themselves, and so on. (A related perspective is to consider the expression of availability as an "input" to the ongoing process though which the various parties *account* for their own actions and for the actions of others [3].) Recent ethnomethodological analyses of mobile phone conversations, such as Laurier's study of mobile white-collar workers [7] and Weilenmann's study of Swedish youth [13], provide updated examples of this phenomenon from today's "wireless world." As Laurier observes, a particular way of communicating "I'm still on the train" that might be interpreted as a simple status update can actually be an important act of preemptive, long-distance face-work [4]. Practices of "getting the right message across" seem important in certain situations, but also seem difficult to accomplish using the simplest representations of availability; this suggests new areas for augmenting such representations.

Considering representations from these perspectives can also help to clarify certain design problems whose existence seems obvious from vague intuitions. For example, a smart availability mechanism that operates as a black-box seems intuitively undesirable because "we don't know what it is doing." Clearly, one problem with such a mechanism is that it takes away contactees' ability to explicitly control the accounts provided to others for their activities and to estimate the accounts that others have constructed. As another example, consider the white-box idea of providing multiple forms of primitive context information (e.g., location, or whether one is alone or not) so that contactors can interpret them and come up with their own formulations of availability [11]. This mechanism is much more easily understood, but we immediately see that the contactee has again lost control over the facts (i.e., context) from which others construct their accounts.

**Determination of availability is a negotiation**
People's willingness to make themselves available to others, and correspondingly their choice of how to represent their current availability to others, depends on a *jointly evolving understanding* of the current situations and needs of all parties. As a phone call proceeds, additional information typically becomes available to the various parties and each continues to assess whether the interaction should continue. An interaction might continue as planned by the contactor, but it might also close early. Closing may be initiated by the contactee, but it can also be initiated by a socially-aware contactor. For example, one can infer from prosody (e.g., tone of voice, or inordinately slow or fast speech) that a contactee is reluctant to continue. As a result, consideration needs to be made of the ways in which an availability system interacts with its associated communication systems to feed information forward from contactors as well as backward from contactees.

By "feed forward," we mean methods of pushing information to the contactee. In conventional telephony, callers with urgent business sometimes cram this fact into the start of a call [13] or, finding themselves in a potential call-screening situation, state their business "in the blind" to a hypothetical listener (i.e., without knowing whether or not the callee is actually listening). In systems where users' first conversational turns serve as the "summons," as in instant messaging (IM) [9] or Nextel push-to-talk cellular radios [14], this first turn is always "in the blind" to some degree. More complex negotiation models in this vein are supported by Quiet Calls [10] and Impromptu [12].

To understand what we mean by "feed backward," consider that communication itself provides context. This is sometimes implicit, as in the leakage of background sound through a phone call (which provides context about current surroundings) or a speaker's prosody. This can be explicit, as in the deliberate verbal sharing of otherwise hidden status [13]. Calls.calm [11] enables contactees to provide more resources to potential contactors for determining availability than a typical presence mechanism. Quiet Calls provides a discreet means for a contactee to provide "verbal" backchannel while a contactor feeds information forward through the open phone connection [10].

Since the final decision that an interaction will proceed as desired by the contactor may result after some amount of interaction has actually already occurred, systems designers can fruitfully anticipate that interactions will unfold in this way. The implication is that, in these cases, integration of "availability" tools and "communication" tools may prove useful – integration that not only goes beyond IM and "away" messages, but even beyond that in Awarenex.

**Availability reflects ongoing relationships**
A given interaction between people with existing relationships often forms a "conversation-in-a-series" [2]. Such interactions are often highly dependent on shared context from previous conversations and closing is often initiated with explicit reference to future contact ("See you," "I'll let you know when I find the file").

Given that this is a frequent situation, it would be useful for a tool that supports contact-making to help manage these bits of conversational context. One way to approach this is to provide separate channels for what Pedersen called "continuity information" in Calls.calm [11]. Continuity information can either feed forward from the contactor (e.g., an "urgent reply" status light can convey that the contactor is upgrading the urgency of a recent request) or backward from the contactee (e.g., a contactor-specific "away" message can convey status such as "I still haven't found the file"). This reflects the fact that new developments can arise on either side of the ongoing interaction. Again, tight integration of "availability" tools and "communication" tools can help in problematic situations.

## SOCIAL, MOBILE AUDIO SPACES

We are currently working on a system that provides mobile, lightweight audio communication within small, tightly-knit social groups, such as college-age friends [1]. The design is influenced, in part, by the recognition that the attempt to engage in conversation is sometimes an integral part of the social negotiation of availability. Rather than focusing on mechanisms for determining presence or availability, we assume that such means will be *partially* effective, and focus instead on a complementary and less-studied problem: smoothing the processes of attempted engagement, dis-engagement and re-engagement in various phases of conversation, doing so in a way that recognizes that not all attempts succeed.

We are applying machine learning techniques to the recognition of various forms of conversational engagement, using voice activity detection and prosodic cues rather than speech recognition. For example, we detect the schisming and merging of parallel conversations in group calls by finding patterns of voice activity that correlate with active turn-taking [1]. The system responds to these patterns by enhancing the intelligibility of talk within a given parallel conversation relative to the talk outside of it. As another example, we are currently working on methods to detect prosodic cues of engaging and dis-engaging talk [14]. Based on this information, the system can choose to alter selected properties of the audio channel in a manner that is consistent with the demonstrated behavior. As a hypothetical example, evidence of dis-engagement (e.g., slowing of speech or of turn-taking) may result in a conversion of the channel from full-duplex to half-duplex push-to-talk, further slowing down turn-taking and thereby facilitating dis-engagement.

Clearly, this work does not address the forecasting of presence and availability directly. However, when considered in light of the discussion in the previous section, we believe that it suggests interesting directions for user interfaces for availability-enhanced communication systems. For example, it suggests that one should work to smooth the process of escalating from asynchronous mechanisms (availability status indicators, away messages) to synchronous mechanisms (voice) – that integration in the form of (e.g.) a binary "click to talk" button in IM does not reflect the more continuous process of conversational engagement. Similarly, it suggests additional areas in which tight integration of availability forecasting and communication might be fruitful. For example, in our system, one could use forecasts as prior information in the assessment of a participant's desire for highly engaged talk.

## CONCLUSION

In everyday life, the expression of availability is often an element of a social negotiation which is bound to situational, temporal and relationship contexts. This implies a number of considerations for designers of availability systems. These considerations further suggest some possible directions for improving the integration of availability and presence mechanisms with the communication mechanisms they support.

## ACKNOWLEDGMENTS

Our project's approach has been greatly influenced by our conversation analyst, Peggy Szymanski. (Any errors or misrepresentations are ours alone.)